\documentclass[a4paper, 12pt, twoside]{article} 
\usepackage[top=2.54cm, bottom=2.54cm, left=2.54cm, right=2.54cm]{geometry} 
\usepackage{setspace}
\usepackage{amsmath,amssymb,amsthm}
\usepackage{booktabs,siunitx}
\usepackage{color,graphicx,hyperref}
\usepackage{times} 
\usepackage{lmodern} 
\usepackage{hanging} 
\usepackage[english]{babel}  
\usepackage{upgreek}

\usepackage{authblk}
\usepackage{multirow}
\usepackage{placeins}
\usepackage{fancyhdr}
\usepackage{cleveref}
\usepackage{rotating,tabularx}
\usepackage{fancyhdr} 
\usepackage{algorithm}
\usepackage{algpseudocode}

\usepackage[super,sort&compress]{natbib}
\setcitestyle{comma,numbers,super,open={},close={}} 

\newcommand{\be}{\begin{equation}}
\newcommand{\ee}{\end{equation}}
\newcommand{\beqa}{\begin{eqnarray*}}
\newcommand{\eeqa}{\end{eqnarray*}}
\newcommand{\beqn}{\begin{eqnarray}}
\newcommand{\eeqn}{\end{eqnarray}}
\newcommand{\ba}{\begin{array}}                                                                                                                                                 \newcommand{\ea}{\end{array}}
\newcommand{\bc}{\begin{center}}
\newcommand{\ec}{\end{center}}
\newcommand{\btab}{\begin{tabular}}
\newcommand{\etab}{\end{tabular}}

\newcommand{\mb}{\makebox}

\newcommand{\lb}{\label}

\newcommand{\nn}{\nonumber}

\newcommand{\argmax}{\mb{\rm arg max}}

\def \nn{\nonumber}

\begin{document}

\begin{center} 
{\fontsize{16}{24} \selectfont 
{\bf Efficient Integration of Aggregate Data and Individual Participant Data in One-Way Mixed Models} } \\[6mm] 
\normalsize
{\bf Neha Agarwala \textsuperscript{1} and Junyong Park \textsuperscript{2} and Anindya Roy\textsuperscript{1}} \\
{\it \textsuperscript{1}Department of Mathematics and Statistics, \\
University of Maryland, Baltimore County, Baltimore, MD, USA} \\[1mm] 
{\it \textsuperscript{2}Department of Statistics, \\
Seoul National University, Seoul, South Korea} \\[6mm] 
\end{center}
\vspace{-8mm}

\noindent\rule{16.5cm}{0.8pt}\vspace{-3mm}

\noindent{\bf Abstract}\vspace{-1mm}

Often both Aggregate Data (AD) studies and Individual Participant Data (IPD) 
studies are available for specific treatments. Combining these two sources of data could
improve the overall meta-analytic estimates of treatment effects. Moreover, often for some studies with AD, the associated IPD maybe available, albeit at some extra effort or cost to the analyst.
We propose a method for combining treatment effects across trials when the response is from the exponential family of distribution and hence a generalized linear model structure can be used.
We consider the case when treatment effects are fixed and common across studies. Using the proposed combination method, we evaluate the wisdom of choosing AD when IPD is available by studying the relative efficiency of analyzing all IPD studies versus combining various percentages of AD and IPD studies. For many different models, design constraints under which the AD estimators are the IPD estimators, and hence fully efficient, are known. For such models we advocate a selection procedure that chooses AD studies over IPD studies in a manner that force least departure from design constraints and hence ensures an efficient combined AD and IPD estimator.

\setlength\parindent{1cm}


\noindent\rule{16.5cm}{1.0pt} \\

\setcounter{equation}{0}

\section{Introduction}

Meta-analysis of individual participant data (IPD) is the gold standard statistical approach in systemic reviews of randomized clinical trials. \cite{chalmers_cochrane_1993} However, if IPD studies systematically differ from studies without access to IPD, synthesizing information solely based on IPD studies may lead to data availability bias or reviewer selection bias. \cite{ahmed2012assessment} In many cases, IPD may not always be publicly available and access to IPD  may be restricted due to limited resources. If only summary  data are available on quantities of interest, meta-analysis of aggregate data (AD) approaches are the only recourse an investigator has for combining information about variables of interest across different studies. In addition to AD studies, if some IPD studies are available, combining these two levels of data could improve the overall meta-analysis estimates, compared to utilizing AD studies alone. \cite{jones2009meta,cooper2009relative} Substitution of AD with IPD is generally advocated `\textit{despite the extra cost, time and complexity required to obtain and manage raw data}'. \cite{riley2010commentary} 
However, in certain situations the meta-analysis based exclusively on aggregate data (AD-MA) may yield estimators that are comparable to those based solely on individual participant data (IPD-MA) and hence substitution of AD with IPD may not be worthwhile due to the extra cost.

In this paper, we discuss how to aggregate information from IPD studies and AD studies  for a generic model and obtain a combined estimate for the parameter of interest that is cost effective yet efficient. We denote the combined estimator based on individual participant data and aggregate data meta-analysis as IPD-AD-MA. We explore many such combinations and investigate trade-offs between efficiency gains from substituting AD with IPD versus the cost of obtaining IPD studies when it is not easily available. The focus of the paper is to analyze whether it matters which combination of IPD and AD are selected in obtaining the IPD-AD-MA estimator. We then propose a novel selection algorithm for selecting which AD to substitute with IPD for maximum efficiency gain. 

One common application of combining studies is testing the effectiveness of new treatments or medical interventions in randomized clinical trials (RCTs). Typically, RCTs tend to differ in study design and are often conducted across different centers, which may result in conflicting evidence. Thus, meta-analysis approaches have become the norm to integrate the results by borrowing power across different trials and provide an all-inclusive conclusion. This paper aims to exploit the design of the trials. Combination of results from IPD and AD studies work well when the model across the studies is simple with a common parameter of interest, e.g. treatment effect, and without disparate covariates. Due to randomization, RCTs on a common treatment provide ideal setup for exploring combination designs of IPD-AD-MA.

Consider $k$ independent studies with the $j$th study resulting in the estimated effect size $\hat{\beta}_j$, an estimate of the population effect size $\beta_j$ for $j =1,\dots,k$. Suppose $\widehat{v(\hat{\beta}_j)}$ is the estimated variance of $\hat{\beta}_j$. When studies are homogeneous i.e, $\beta_j=\beta$ for all $j$, $\beta$ is estimated efficiently by a fixed effects model, using weighted combination of $\hat{\beta}_j$ with the weights taken to be the reciprocal of $\widehat{v(\hat{\beta}_j)}$. When studies are heterogeneous, a random effects model is considered to account for the between study variances, $\tau^2$. The meta-analysis estimate of $\beta$ is a weighted mean of $\widehat{v(\hat{\beta}_j)}$  where the weights for each study is given by the reciprocal of $\widehat{v(\hat{\beta}_j)} + \tau^2$. Some widely used estimators for $\tau^2$ are based on the Cochran's homogeneity test statistic.\cite{dersimonian1986meta,dersimonian2015meta,viechtbauer2010conducting} Several other estimators for the heterogeneity variance are available.\cite{veroniki2016methods,panityakul2013estimating,sidik2007comparison,viechtbauer2005bias,hedges2014statistical,hartung2011statistical,mcculloch2005generalized} Many papers like van Houwelingen \textit{et al} (2002), Ritz \textit{et al} (2008) have investigated the multivariate extension of these meta-analysis models.\cite{van2002advanced,ritz2008multivariate}

The meta-analysis literature is rich with efficiency comparison of AD-MA estimators with the estimators obtained from full data based on different models. For treatment vs. control comparison with continuous outcome, Olkin and Sampson (1998) showed AD-MA estimator is equivalent to IPD-MA estimator if there is no study-by-treatment interactions and variances are constant across trials.\cite{olkin1998comparison} Mathew and Nordstrom (1999) further showed that this equivalence holds even if the error variances are different across trials.\cite{mathew1999equivalence} The performance of IPD-MA estimator has been found to be similar but not identical to AD-MA estimator empirically (e.g. Whitehead, 2002, Ch. 5).\cite{whitehead2002meta} For a more general linear model with fixed treatment and random trial effect, Mathew and Nordstorm, 2010 provided conditions under which AD analysis and IPD analysis coincide.\cite{mathew2010comparison} For all commonly used parametric and semi-parametric models, Lin and Zeng (2010) showed that IPD-MA estimator has no gain in efficiency over AD-MA estimator asymptotically in the context of fixed effects model and also provided the condition for their equality.\cite{lin2010relative} Liu \textit{et al} (2015) introduced a meta-analysis approach for heterogeneous studies by combining the confidence density functions derived from the summary statistics of individual studies.\cite{liu2015multivariate} Doneal \textit{et al} (2015) compared the performance of IPD-MA and AD-MA using different estimation procedures in generalized linear mixed model for binary outcomes.\cite{thomas2017comparison} 

One potential limitation in standard meta-analysis approach is the requirement of a common set of parameters across studies. Different studies often tend to include different sets of covariates. For meta-analysis of IPD, Jackson \textit{et al} proposed a method to estimate the fully adjusted effect across studies with different set of confounders.\cite{fibrinogen2009systematically} Recently, Kundu \textit{et al} developed an extension of meta-analysis method for fixed-effects model to combine information from studies with disparate covariate information.\cite{kundu2019generalized} 

An extensive efficacy analysis of the one-stage and two-stage statistical methods for combining IPD and AD in meta-analysis for continuous outcome was explored by Riley \textit{et al} among others.\cite{riley2008meta1,riley2008meta,riley2007evidence,idris2015study} In the simple situation of a fixed effects model with a single continuous outcome and covariate, Yamaguchi \textit{et al} proposed a method to reconstruct the missing IPD for AD trials by a Bayesian sampling procedure and use the mixture of simulated IPD and collected IPD for an IPD meta-analysis.\cite{yamaguchi2014meta} Over the past decade, meta-analysis methods for mixture data have also been developed for dichotomous outcomes and time-to-event data, some based on reconstruction of IPD.\cite{guyot2012enhanced,riley2010meta,riley2007evidence} Other popular methods for integrating binary data is random-effects mixed treatment comparison (MTC) models and likelihood based approaches.\cite{donegan2013combining,saramago2012mixed,ravva2014linearization,sutton2008meta,jackson2006improving}

In section \ref{section_lmm}, we start with the case of a continuous response following a  linear model and provide the combined treatment effects across trials when the treatment effect is fixed and common across trials while the trial effect is random. Assuming the observations within and between the studies are independent, we investigate the loss of efficiency from using combined estimator with various percentages of AD and IPD studies. When treatments are fixed and trial effects are random, Mathew and Nordstorm (2010) derived the necessary and sufficient condition for the IPD-MA estimator to coincide with AD-MA estimator for a general within trial covariance matrix. \cite{mathew2010comparison} The condition for equality requires that the fraction of observations corresponding to any given treatment to be  same across trials. In practice, it is more likely to have studies with differential allocation to treatments. For such models, we study the relative efficiency of an estimator based on combining IPD and AD studies, denoted by IPD-AD-MA to the IPD-MA estimator under systematic departures from the same allocation proportion condition. We further propose a method to select the IPDs among the available studies so as to get the maximum efficiency in terms of the combined estimator. 

In section \ref{section_glmm}, we propose a method of combining information across IPD and AD studies for a multidimensional parameter in a generalized linear mixed model (GLMM) and study the performance of the combined estimator empirically. This is a more general setup where the covariate or response may be categorical or continuous and the common parameter of interest can be multidimensional. In addition, the random effects may not necessarily be independent. As a special case, we consider a logistic model with a similar setup to the LMM framework. For this case, we derive a relative efficiency expression and use the expression to propose efficient selection of IPD when combining IPD and AD studies for the IPD-AD-MA estimator.  

We use a real data example to illustrate efficient selection of IPD when synthesizing information on the common parameter of interest across IPD and AD studies for a linear mixed model and a logistic model with random study effect. For all our analyses, we assume that all studies or trials are independent, which is a general and common assumption in the meta-analysis literature.
\section{Efficient Aggregation in Linear Mixed Model (LMM)}\label{section_lmm}

Consider that there is one continuous outcome of interest and assume that there are two groups, namely treatment(T) group  and control(C) group for all $k$ independent studies. Let $y_{ji}$ be the response of the $i$th participant in study $j$ and $x_{ji}$ is the corresponding treatment indicator. Suppose further that, $n_{jT}$ and $n_{jC}$ be the number of persons in the treatment group and the control group, respectively, for $j$th study with $n_{jT}+n_{jC}=n_j$.  Suppose, for $k_1$ studies, IPD are available and for the remaining $k-k_1=k_2$ studies, we have access to only AD. Let $S_1$ and $S_2$ denote the set of IPD studies and AD studies, respectively where $S_1 \cup S_2= S$, $S$ being the set with all studies. The model is

\begin{equation}
\setlength{\jot}{10pt}
\begin{aligned}
    y_{ji} = \alpha_j + \beta x_{ji} + \epsilon_{ji} , \quad j =1,...,k,\\
     \epsilon_{ji} \sim N(0,\sigma_j^2),  \quad \alpha_j \sim N(\alpha,\sigma_{\alpha}^2), \\
\end{aligned}
\label{lmm_model}
\end{equation}
where $\alpha_j$ and $\epsilon_{ji}$ are assumed independent. Our parameter of interest is the common treatment effect $\beta$.
\subsection{Aggregation}
For model \eqref{lmm_model}, $\alpha$ acts as a nuisance parameter common across studies. Thus, the AD-MA estimator and the IPD-MA estimator does not necessarily coincide although the parameter of interest is common across studies (Mathew and Nordstorm 2010).\cite{mathew2010comparison} In fact, the IPD-MA estimator, which is the Best Linear Unbiased Estimator (BLUE), is more efficient than the AD-MA estimator for any finite and fixed $k$ and $n_j$. \cite{mathew2010comparison} The two estimators coincide if and only if the vectors $(n_{jT}/n_j,n_{j2}/n_j)$ are all equal for $j=1,...,k$. Asymptotically, the AD-MA estimator has the same efficiency as the IPD-MA estimator.

For the $k_1$ studies with access to IPD, $\textbf{y}_j$ is normal with mean and covariance given by
\begin{equation*}
\setlength{\jot}{10pt}
\begin{aligned}
    E(\boldsymbol{y_j}|X_j) &= \alpha \boldsymbol{1_{n_{j}}} + \beta X_j \\
    Cov(\boldsymbol{y_j}|X_j) &= H_j = \sigma_{\alpha}^2 \boldsymbol{1_{n_{j}}}\boldsymbol{1_{n_{j}}^T} + \sigma_j^2 I_{n_{j}}.\\
\end{aligned}
\end{equation*}
For the $k_2$ AD studies, the maximum likelihood estimates (MLE), $\hat{\beta_j}$, and their estimated variances, $\widehat{v(\hat{\beta}_j)}$ are available. The model for AD study is
\begin{equation*}
\setlength{\jot}{10pt}
\begin{aligned}
 \hat{\beta}_j \sim N(\beta,\widehat{v(\hat{\beta}_j)}), \quad j =1,\dots,k_2 \\
 v(\hat{\beta}_{j}) = {\big(n_{j} \pi_{j} (1-\pi_j)\big)}^{-1}{\sigma_j^2}\\
\end{aligned}
\end{equation*}
where $\pi_j = n_{jT}/n_j$ is the proportion of treatment in study $j$. 

Integration of IPD and AD uses the standard weighted combination approach with weights being inversely proportional to the variance, where the variance for AD part is simply the variance of MLE whereas for the IPD, it is the variance-covariance matrix of the marginal distribution of the data. With the above model, the combined estimator of $\boldsymbol{\theta} = (\alpha,\beta)^{\prime}$ and the variance of the combined estimator are
\beqn
 \hat{\boldsymbol{\theta}}_{IPD-AD-MA} &=& (U^T\Sigma^{-1} U)^{-1} U^T\Sigma^{-1} Y^*, \nn \\
Cov( \hat{\boldsymbol{\theta}}_{IPD-AD-MA} ) &=& (U^T\Sigma^{-1} U)^{-1},
\lb{combined_est}
\eeqn
where $Y^* = (\textbf{y}_1^{\prime},\dots,\textbf{y}_{k_1}^{\prime}, \hat{\beta}_1,\dots,\hat{\beta}_{k_2})^{\prime}$ and 
\begin{equation*}
\setlength{\jot}{10pt}
\begin{aligned}
U = \begin{pmatrix}
           \boldsymbol{1_{n_{1}}} & X_1 \\
           \dots & \dots \\
            \boldsymbol{1_{n_{k_1}}} & X_{k_1} \\
           0 & \boldsymbol{1_{k_2}}\\
         \end{pmatrix},\;
\Sigma = \begin{pmatrix}
           \Sigma_1 & 0 \\
           0 & \Sigma_2\\
         \end{pmatrix},\; 
\Sigma_1 &= \begin{pmatrix}
        H_1\\
        &\ddots\\
        &&H_{k_1}\\
        \end{pmatrix},\;
\Sigma_2 = \begin{pmatrix}
        v(\hat{\beta}_{1})\\
        &\ddots\\
        &&v(\hat{\beta}_{k_{2}})\\
        \end{pmatrix}.\\
\end{aligned}
\end{equation*}

Since the estimator in (\ref{combined_est}) is unbiased, for analyzing efficiency of the estimator, we obtain the expression for its variance. To derive the variance of the combined estimator for the treatment effect, we assume $\sigma_j^2$ and $\sigma_{\alpha}^2$ are known. However, both the variance components are estimated from the data in section \ref{lmm_sim_sensitivity} to reflect practical considerations. The variance is given by
\begin{equation}\label{lmm_var}
\setlength{\jot}{10pt}
\begin{aligned}
v (\hat{\beta}_{IPD-AD-MA})  ={\bigg[\sum_{j \in S_{1}} \frac{n_j \pi_{j} (1+n_j (1-\pi_{j})b_j)}{\sigma_j^2 a_j} + \sum_{j \in S_{2}} \frac{n_j \pi_{j} (1-\pi_{j})}{\sigma_j^2} - \frac{ (\sum_{j \in S_{1}} \frac{n_j \pi_{j}} {\sigma_j^2 a_j})^2}{\sum_{j \in S_{1}} \frac{n_j} {\sigma_j^2 a_j}}  \bigg]}^{-1}
\end{aligned}
\end{equation}
where $a_j=1+n_j b_j, \; b_j=\frac{\sigma^2_{\alpha}}{\sigma_j^2} \; \& \; \pi_j = \frac{n_{jT}}{n_j}$. 

The gold standard for the combined analysis is having access to all IPD studies. Thus, we compare the efficiency of $\hat{\beta}_{IPD-AD-MA}$ with the ``best" estimator, which is the maximum likelihood estimate (MLE) based on IPD from all studies. The variance of the ``best" estimator of $\beta$, also known as the minimum variance unbiased estimator is 
\begin{equation*}
\setlength{\jot}{10pt}
\begin{aligned}
v (\hat{\beta}_{IPD-MA})  ={\bigg[\sum_{j=1}^{k} \frac{n_j \pi_{j} (1+n_j (1-\pi_{j})b_j)}{\sigma_j^2 a_j} - \frac{ \big(\sum_{j=1}^{k} \frac{n_j \pi_{j}} {\sigma_j^2 a_j}\big)^2}{\sum_{j=1}^{k} \frac{n_j} {\sigma_j^2 a_j}}  \bigg]}^{-1}.
\end{aligned}
\end{equation*}

The  variance of the combined  estimator of $\beta$, assuming only AD results are available for all $k$ studies, is 
\begin{equation*}
\setlength{\jot}{10pt}
\begin{aligned}
v (\hat{\beta}_{AD-MA})  ={\bigg[ \sum_{j=1}^{k} \frac{n_j \pi_{j} (1-\pi_{j})}{\sigma_j^2} \bigg]}^{-1}.
\end{aligned}
\end{equation*}

For the balanced homoscedastic case, when $n_j=n$ and $\sigma_j^2 = \sigma^2$,  the variance expression for the combined estimator simplifies to
\begin{equation*}
\setlength{\jot}{10pt}
\begin{aligned}
v (\hat{\beta}_{IPD-AD-MA})  =\frac{\sigma^2}{n}{\bigg[\frac{n b}{a} \sum_{j \in S_{1}} \pi_{j} (1-\pi_{j}) + \sum_{j \in S_{2}} \pi_{i} (1-\pi_{i}) + \frac{1}{ak_1} \Big(\sum_{j \in S_{1}} \pi_{j}\Big) \Big(\sum_{j \in S_{1}} (1-\pi_{j})\Big) \bigg]}^{-1}
\end{aligned}
\end{equation*}
where $a=1+nb, \; b=\frac{\sigma^2_{\alpha}}{\sigma^2}$. Similarly, the variance expression for the all IPD estimator reduces to
\begin{equation*}
\setlength{\jot}{10pt}
\begin{aligned}
v (\hat{\beta}_{IPD-MA})  &= \sigma^2{\bigg[\frac{n^2 b}{a} \sum_{j=1}^{k} \pi_{j} (1-\pi_{j}) + \frac{n}{ak} \Big(\sum_{j=1}^{k} \pi_{j}\Big) \Big(\sum_{j=1}^{k} (1-\pi_{j})\Big) \bigg]}^{-1}.
\end{aligned}
\end{equation*}

The relative efficiency of the combined estimator with respect to the estimator with all IPD studies is
$RE (\hat{\beta}_{IPD-AD-MA}) = \frac{v(\hat{\beta}_{IPD-MA})}{v(\hat{\beta}_{IPD-AD-MA})}$ which for the simple case when $n_j=n$ and $\sigma_j^2 = \sigma^2$ is given by
\begin{equation}
          \frac{\bigg[\frac{n^2 b}{a} \sum_{j \in S_1} \pi_{j}(1-\pi_{j}) + n \sum_{j \in S_2} \pi_{j}(1-\pi_{j}) + \frac{n}{ak_1} \Big(\sum_{j \in S_1} \pi_{j}\Big) \Big(\sum_{j \in S_1} (1-\pi_{j})\Big) \bigg]}{{\bigg[\frac{n^2 b}{a} \sum_{j=1}^{k} \pi_{j} (1-\pi_{j}) + \frac{n}{ak} \Big(\sum_{j=1}^{k} \pi_{j}\Big) \Big(\sum_{j=1}^{k} (1-\pi_{j})\Big) \bigg]}}.
\end{equation}
We will use the above expression for relative efficiency for simulation purposes in section \ref{lmm_sim}.

\subsection{Selection}\label{selection}

The main question that we address in this paper is whether, while substituting IPD studies with the corresponding AD results or vise versa, it matters which  studies are selected. If yes, how can we minimize the loss in efficiency occurring from using AD results instead of the IPD results for a study? 

The variance of the combined estimator in equation (\ref{lmm_var}) can be simplified to 

\begin{equation}\label{var_lmm_simple}
\setlength{\jot}{10pt}
\begin{aligned}
v (\hat{\beta}_{IPD-AD-MA})  ={\bigg[\sum_{j \in S_{1}} \frac{n_j}{\sigma_j^2 a_j} (\pi_{j}-\tilde{\pi}_{S_1})^2 + \sum_{j =1}^{k} \frac{n_j \pi_{j} (1-\pi_{j})}{\sigma_j^2}   \bigg]}^{-1}
\end{aligned}
\end{equation}
where $\tilde{\pi}_{S_1} = \frac{\sum_{j \in S_{1}} {\frac{\pi_j n_j}{\sigma_j^2 a_j}}} {\sum_{j \in S_{1}} {\frac{n_j}{\sigma_j^2 a_j}}}$.

The  expression for variance in equation (\ref{var_lmm_simple}) does not involve the response $\boldsymbol{y}$ and depends only on $\pi_j$, $n_j$ and $\sigma_j$. This gives us a way of selecting $k_1$ IPD studies among the $k$ studies to optimize the efficiency of combined estimator. In other words, selection of IPD studies depends on choosing the best subset $S_1$ that maximizes $\sum_{j \in S_{1}} \frac{n_j}{\sigma_j^2 a_j} (\pi_{j}-\tilde{\pi}_{S_1})^2$. Essentially this is a combinatorial optimization problem which can be formulated as follows:\\

\noindent{\bf Selection Problem:} $\,$Given $k > 1$ and $1 \leq k_1 \leq k$, and $\mathcal{P}_k$, the power set of $ \{1,\dots,k\}$, find
\begin{equation}
  A_{opt} = \underset{A \in \mathcal{P}_k : |A| = k_1}\argmax\, \sum_{j \in A} v_j(u_j-\Bar{u}_A)^2  
\label{opt_selection}
\end{equation}
where $(u_1,v_1),\dots,(u_k,v_k)$ with $u_j>0$, $v_j>0$, are prespecified constants with $\Bar{u}_A=\frac{\sum_{j \in A} v_j u_j}{\sum_{j \in A} u_j}$ for any $A \in \mathcal{P}_k$.\\

For the balanced homoscedastic case, the variance further reduces to
\begin{equation*}
\setlength{\jot}{10pt}
\begin{aligned}
v (\hat{\beta}_{IPD-AD-MA})  ={\frac{\sigma^2}{n} \bigg[\sum_{j \in S_{1}} \frac{1}{a} (\pi_{j}-\tilde{\pi}_{S_1})^2 + \sum_{j \in S} \pi_{j} (1-\pi_{j})   \bigg]}^{-1}.
\end{aligned}
\end{equation*}
For this special case, an exact algorithm exists for finding the optimum set, $A_{opt}$ of IPD studies. Firstly we arrange the studies in increasing order of their $\pi_j$ values. We include studies with extreme $\pi_j$ values in the IPD set, alternating between the two ends to choose a total of $k_1$ IPD studies. 

This is consistent with the result from Mathew and Nordstorm (2010) where the IPD-MA estimator coincides with AD-MA estimator when the fraction of observations corresponding to any given treatment is same across trials in a linear model with fixed treatments and random trial effects. When combining IPD and AD studies, we propose allocating studies with similar proportion of treatments to the AD set and studies with widely varying proportion of treatments to the IPD set.

\begin{algorithm}[H]
\caption{Sequential Selection Algorithm: SSA}\label{seq_algm}
\begin{algorithmic}
\Require $k_1 \geq 2$
\State $\mathcal{S} \leftarrow  1:k$
\State $(m, n) \leftarrow \underset{p,q \in \mathcal{S}} {\arg \max}\; (u[p] - u[q])^2/[1/v[p] + 1/v[q])$
\State $A_{SSA} \leftarrow \{m, n\}$
\State $B \leftarrow \mathcal{S}\backslash A_{SSA}$
\State $D \leftarrow v[m] + v[n]$
\State $M \leftarrow (v[m]*u[m] + v[n]*u[n])/D$
\While{$|A_{SSA}| \neq k_1$}
\State $ l \leftarrow  \underset{r \in B} {\arg \max}\; v[r]*(u[r] - M)^2/(D + v[r])$
\State $ A_{SSA} \leftarrow A_{SSA} \bigcup \{l\}$
\State $ B \leftarrow B\backslash \{l\}$
\State $ M \leftarrow (D*M + v[l]*u[l])/(D + v[l])$
\State $ D \leftarrow D + v[l]$
\EndWhile
\end{algorithmic}
\label{algo}
\end{algorithm}

For a  more general case when $n_j$ or $\sigma_j$ may not be equal for all $j$, we propose an approximate sequential algorithm for the selection problem stated in (\ref{opt_selection}). The sequential selection algorithm, SSA is a forward selection algorithm that involves choosing the first two studies so that the objective function is maximized. Then, we select the next $k_1-2$ studies sequentially using the weighted mean based on the preceding selected studies. It sidesteps the computationally intensive task of calculating the weighted mean for every subset $S_1$ with cardinality $k_1$. Two vectors, $u$ and $v$ are fed as an input to the SSA algorithm and the output is the set $A_{SSA}$ of $k_1$ studies. The performance of the algorithm is assessed using simulation studies and the results are presented in section \ref{lmm_sim2}. 
\subsection{Simulation}\label{lmm_sim}

Assuming the variance components are known without uncertainty, the variance expression in (\ref{var_lmm_simple}) does not involve any data and hence the results in the following sections do not involve any simulation. The discussion on estimation of the variance components and its sensitivity to the selection algorithm is covered in section \ref{lmm_sim_sensitivity}. 

For linear mixed effects model, we present two scenarios to show the importance of selecting appropriate studies for constructing the combined estimator. For simplicity, we consider both $k$ and $n$ to be small and equal to $10$ and $\sigma^2 = 2.5, \sigma_{\alpha}^2 = 0.025$.

\subsubsection{Uniform distribution of proportion of treatment}\label{lmm_sim_case1}

For this scenario, we use the uniform distribution to randomly generate $k=10$ proportion of treatment over the interval $[0,1]$ as provided in Table \ref{prop1}:

\begin{table}[ht]
  \centering
  \caption{Uniform distribution of proportion of treatment across studies}
  \vspace*{1mm}
  \begin{tabular}{|c|cccccccccc|}
  \hline
  \textbf{study}& 1 & 2 & 3 & 4 &  5 &  6 &  7 &  8 & 9	&  10\\
  \hline
  \textbf{proportion}& 0.1 & 0.2 & 0.3 & 0.3 & 0.3 & 0.5 & 0.6 & 0.6 & 0.8	& 0.8\\
  \hline
  \end{tabular}
  \label{prop1}
  \end{table}

\begin{figure}[H]
    \centering
    \includegraphics[width=5in]{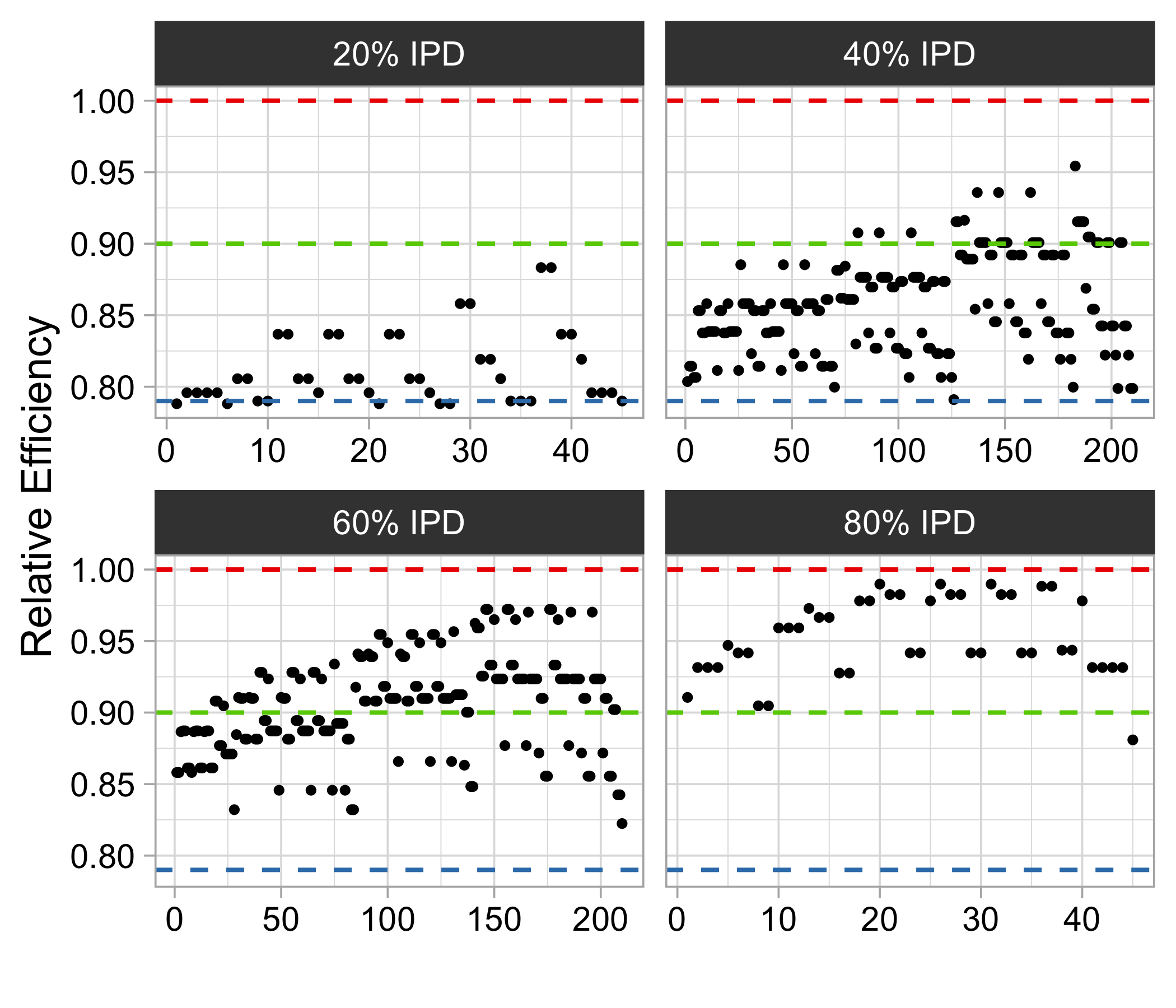}
    \caption{The relative efficiency of all 45 possible combinations for each of 20\% IPD and 80\% IPD, 210 possible combinations for each of 40\% IPD and 60\% IPD are plotted. Red dashed line represents the RE for IPD-MA estimator which is 1, blue dashed line is RE for AD-MA estimator which is 0.79 and the green dashed line represents the desired RE, say 0.9, for example.}
    \label{fig1}
\end{figure}

Figure \ref{fig1} shows that if the desired relative efficiency is 0.9, we could achieve that with many possible combinations of 60\% IPD and 40\% AD studies. However 60\% IPD may not be cost effective. But if we choose the right combination of 40\% IPD and 60\% AD, we can have a RE of 0.956 in this case. One such combination which maximizes the RE for 40\% IPD is study no 1, 2, 9 and 10 for IPD with treatment proportion 0.1, 0.2, 0.8 and 0.8, respectively, and AD for the rest of the studies.

\begin{figure}
    \centering
    \includegraphics[width = 5in]{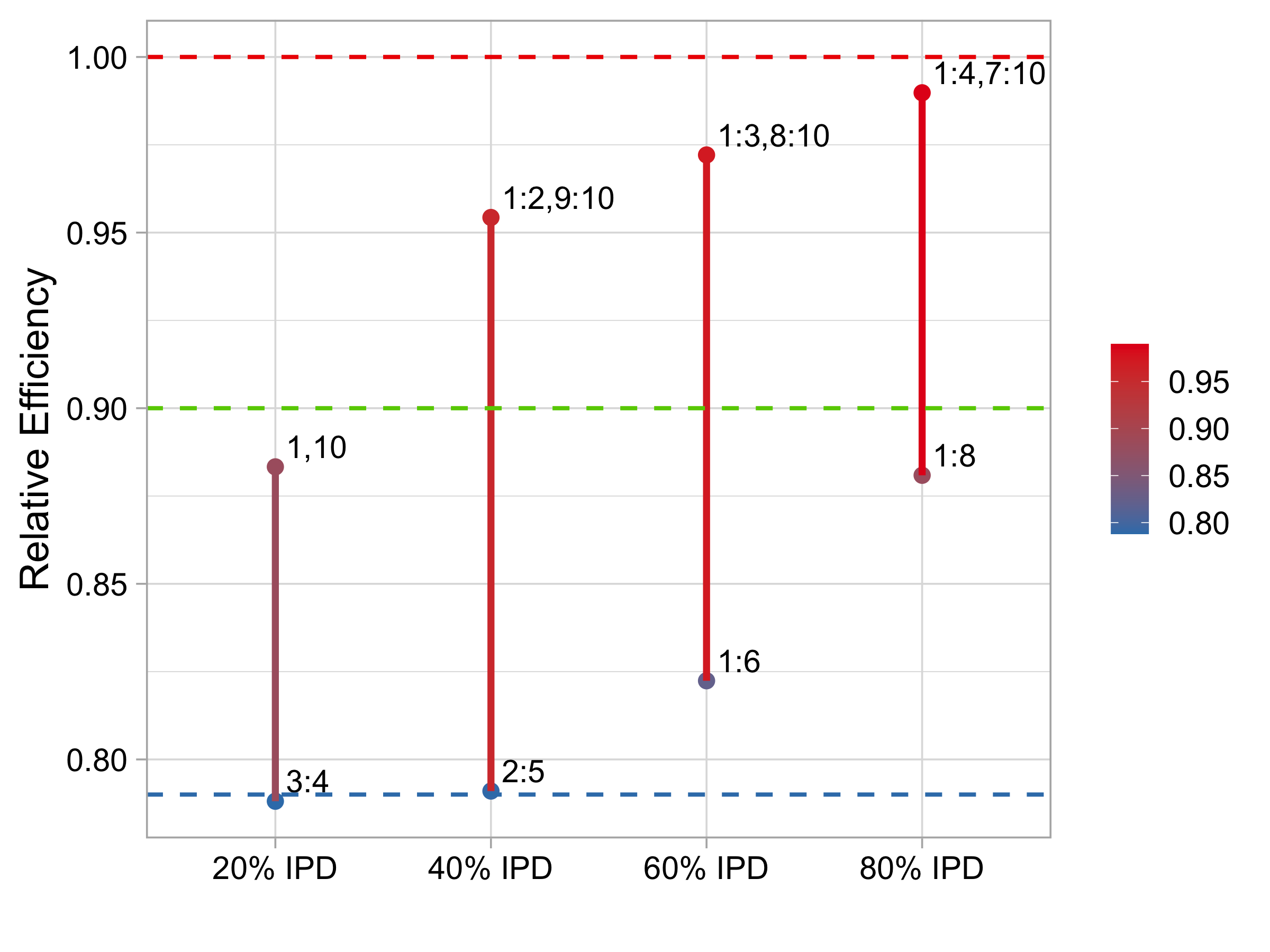}
    \caption{Plot showing the maximum and minimum relative efficiency for each percentage of IPD studies and the numbers at the top and bottom of the line plot indicate the IPD studies which achieves the maximum and minimum relative efficiency, respectively.}
    \label{fig2}
\end{figure}

Figure \ref{fig2} shows the maximum and minimum relative efficiency that can be achieved for different combinations of the IPD and AD studies. In the figure, the numbers at the top and bottom of the line plot indicate the IPD studies which achieve the maximum and minimum relative efficiency, respectively. For example, among all combinations with 20\% IPD and 80\% AD, the combination which yields maximum relative efficiency includes study 1 and 10 for the IPD part and rest of the studies for the AD part. The minimum efficiency is achieved with the combination of study 3 and 4 for the IPD part and the rest for the AD part.

\subsubsection{Bathtub distribution of proportion of treatment}\label{lmm_sim_case2}

\begin{figure}
    \centering
    \includegraphics[width = 5in]{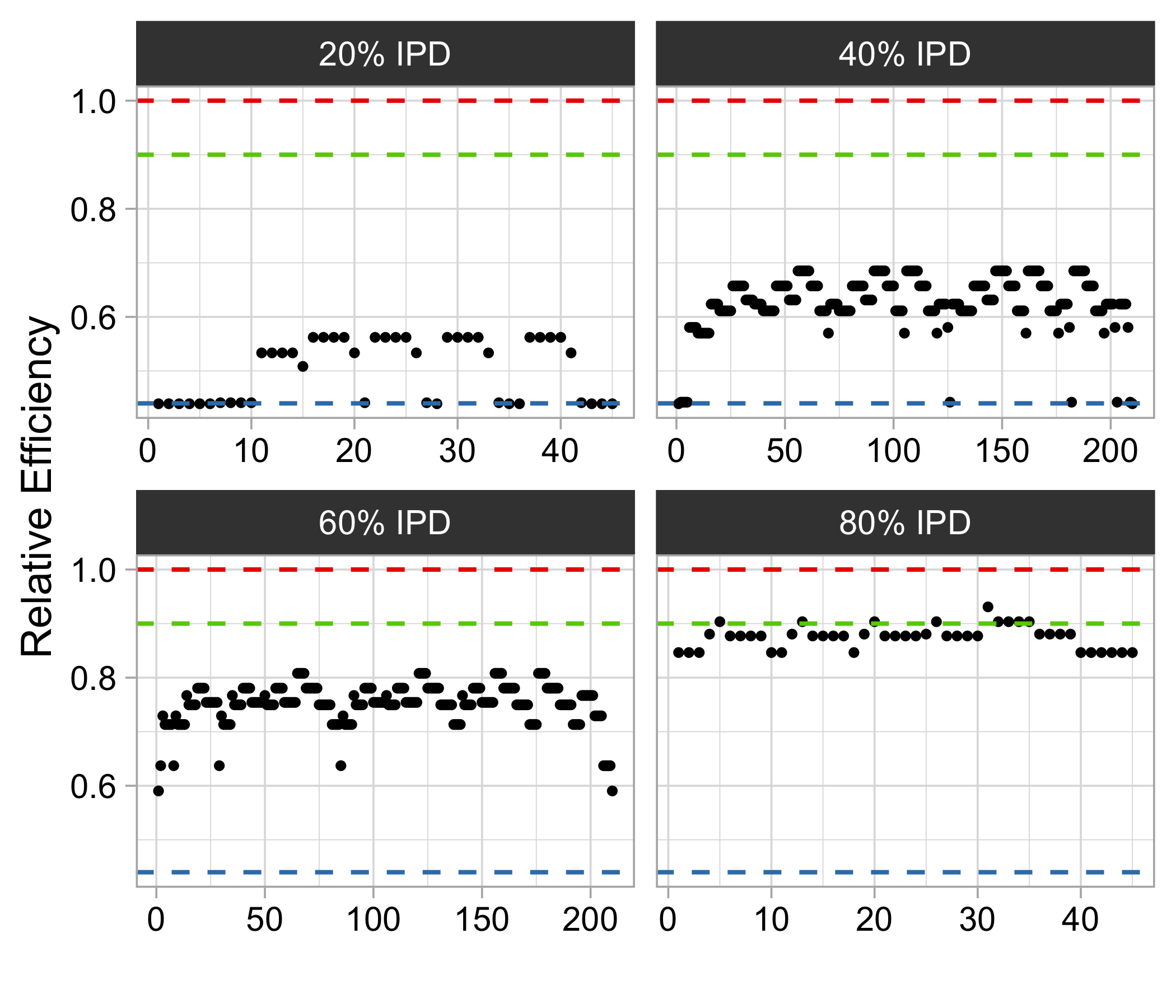}
    \caption{The relative efficiency of all 45 possible combinations for each of 20\% IPD and 80\% IPD, 210 possible combinations for each of 40\% IPD and 60\% IPD are plotted. Red dashed line represents the RE for IPD-MA estimator which is 1, blue dashed line is RE for AD-MA estimator which is 0.79 and the green dashed line represents the desired RE, say 0.9, for example.}
    \label{fig3}
\end{figure}

An example of unbalanced proportion of treatment is given in Table \ref{prop2}. The importance of selecting the best subset of IPD studies for the combined estimate is more clearly seen in this case. The AD-MA estimator has a relative efficiency of $0.44$ compared to the IPD-MA estimator. From Figure \ref{fig3}, we see that it is hard to obtain a relative efficiency of 0.9 even with 80\% IPD and 20\% AD.

\begin{table}[ht]
  \centering
  \caption{Unbalanced distribution of proportion of treatment across studies}
  \vspace*{1mm}
  \begin{tabular}{|c|cccccccccc|}
  \hline
  \textbf{study}& 1 & 2 & 3 & 4 &  5 &  6 &  7 &  8 & 9	&  10\\
  \hline
  \textbf{proportion}&0.1 & 0.1 & 0.1 & 0.1 & 0.2 & 0.8 & 0.9 & 0.9 & 0.9 & 0.9\\
  \hline
  \end{tabular}
  \label{prop2}
  \end{table}

\begin{figure}
    \centering
    \includegraphics[width = 5in]{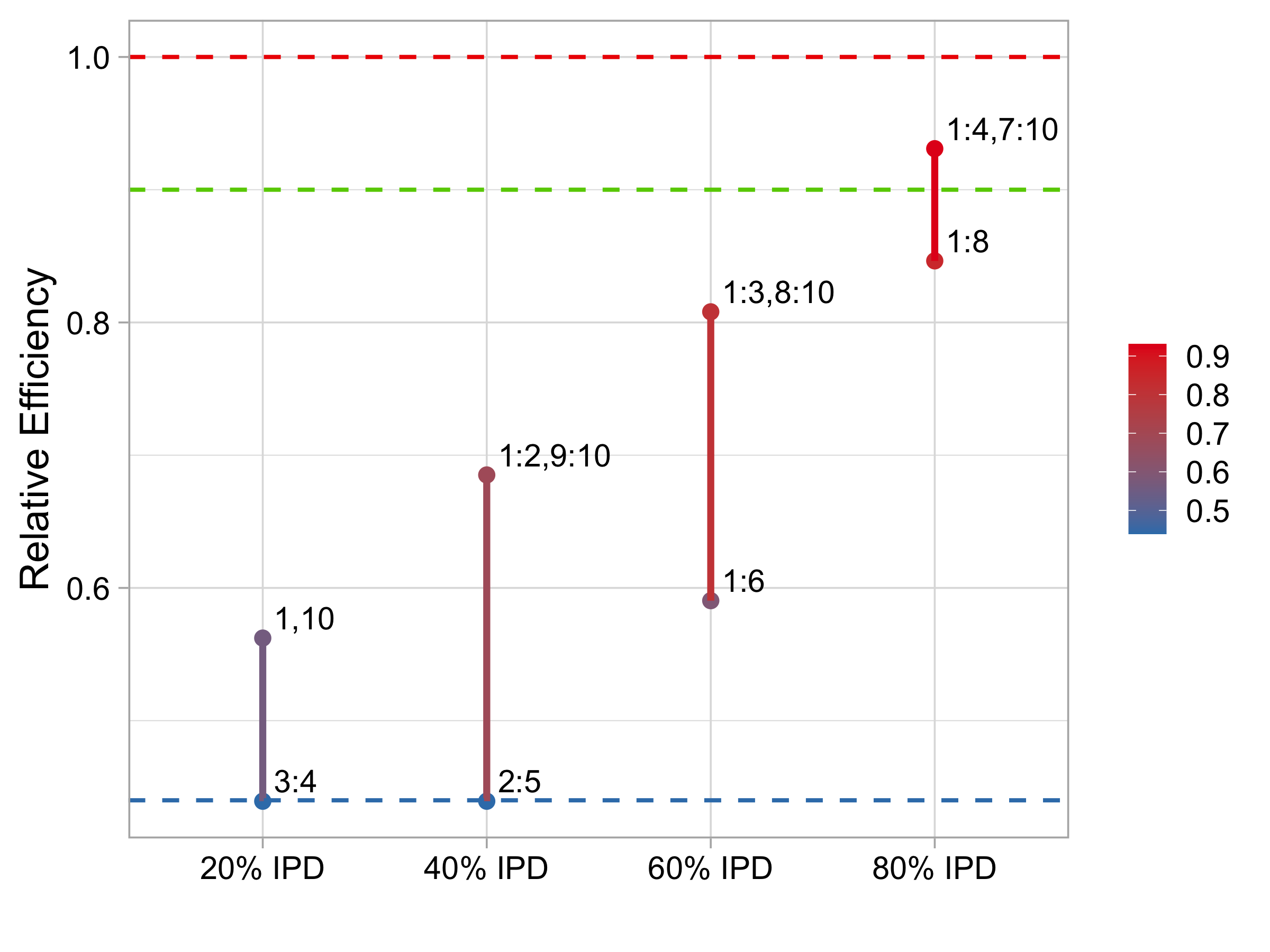}
    \caption{Plot showing the maximum and minimum relative efficiency for each percentage of IPD studies and the numbers at the top and bottom of the line plot indicate the IPD studies which achieves the maximum and minimum relative efficiency, respectively.}
    \label{fig4}
\end{figure}

The situation can worsen for severely unbalanced distribution of proportion of treatment. However, the loss in efficiency is an issue only with fixed small sample size studies. Asymptotically, the AD-MA combined estimator are quite efficient compared to IPD-MA combined estimator. Hence, the selection of IPD in data integration is useful for when combining treatment effect across randomized clinical trials (RCTs) with small cohort size. In such cases, one can obtain maximum efficiency of the combined estimator for a given number of IPD studies using the selection algorithm. 

\subsubsection{Selection of IPD studies through sequential algorithm}\label{lmm_sim2}
To assess the performance of sequential algorithm when $n_j$ or $\sigma_j$ are not equal, we consider $k=30$ studies each with $n=10$ and $\sigma_{\alpha}^2 = 0.025$. We simulate the $\sigma_j$'s and $\pi_j$'s as following:

\begin{equation*}
\setlength{\jot}{10pt}
\begin{aligned}
\pi &\sim 0.5 \ \text{Beta}(\alpha=2,\beta=9)+0.5 \ \text{Beta}(\alpha=9,\beta=2),\\
\sigma^2 &\sim \text{Inv-Gamma}(\alpha=2,\beta=5).
\end{aligned}
\end{equation*}
\\
For $k_1=2,\dots,10$, we first obtain the optimal set with $k_1$ IPD studies for which the variance is minimized and then record the number of times the optimal set matches with the set obtained through the sequential algorithm. Table \ref{sim_lmm_sequential} presents the count of the matches for 100 simulations and the mean ratio of variance for the set from sequential algorithm to the variance of the optimal set. For larger $k_1$, the number of combinations is large and hence finding the optimal set can be time consuming and computationally expensive. The sequential algorithm, on the other hand, is quite fast and performs relatively well. 

\begin{table}[!htbp]
\centering
\caption{Count of matches for a set of size $k_1$ out of $k=10$ studies}
\vspace*{1mm}
\begin{tabular}{|c|c|c|c|c|c|c|c|c|c|}
\hline
& \multicolumn{9}{c|}{$k_1$}\\
\cline{2-10}
\multicolumn{1}{|l|}{\textbf{Count Match}} & {$10$} & {$9$} & {$8$} & {$7$} & {$6$} & {$5$} & {$4$} & {$3$} & {$2$} \\ 
\hline
{0} 		  & 0 	& 0  & 0  & 0  & 0  & 0   & 0 	& 0  & 0 	     \\  
{1}        & 0 	& 0  & 0  & 0  & 0  & 0   & 0 	& 2  & 0       \\  
{2}        & 0 	& 0  & 0  & 0  & 0  & 0   & 2 	& 8  & 100    \\  
{3}        & 0 	& 0  & 0  & 0  & 0  & 3   & 15 	& 90 &          \\  
{4}        & 0 	& 0  & 0  & 0  & 0  & 14 & 83  &     &          \\  
{5}        & 0 	& 0  & 0  & 0  & 7  & 83 &      &     &          \\ 
{6}        & 0 	& 0  & 0  & 3  & 93&      &      &     &          \\  
{7}        & 0 	& 0  & 6  & 97&     &      &      &     &         \\  
{8}        & 0 	& 6  & 94&     &     &     &      &      &         \\  
{9}        & 9    & 94&     &     &     &     &      &     &          \\  
{10}      & 91	&    &     &     &     &     &      &      &         \\  
\hline
{Mean Ratio} & 1 & 1 & 1 & 1 & 1 & 1.001 & 1.001 & 1.001 & 1 \\ 
\hline                       
\end{tabular}
\label{sim_lmm_sequential}
\end{table}

\subsubsection{Variance component estimation}\label{lmm_sim_sensitivity}

In this section, we discuss about the estimation of the variance components $\sigma^2_j$ and $\sigma^2_{\alpha}$ and its sensitivity to the sequential selection algorithm. When $\sigma^2_j \neq \sigma^2_l$  for some $j,l$,  $\sigma^2_j$ is estimated using the estimated variance of the treatment effect and the number of participants for the two groups in each study. One can also test for homogeneity of study variances and obtain a pooled estimate of $\sigma^2$ with the assumption $\sigma^2_j=\sigma^2$ $\forall j$. Furthermore, $\sigma_{\alpha}^2$ is estimated using some pilot IPD studies, that are accessible without any difficulty or effort. 

To study the sensitivity of the variance component estimation to the selection algorithm, two different scenarios are considered: $\sigma^2_j=\sigma^2$ $\forall j$ and $\sigma^2_j \neq \sigma^2_l$  for some $j,l$. For each scenario, we generate $10000$ data sets from the linear mixed effects model in (\ref{lmm_model}) where $\alpha = 0.5, \beta = 1.5, \sigma^2_{\alpha} = 0.025$ with $k = 10$ studies with $n_j = 50$ participants and proportion of treatment as provided in table \ref{prop2}. The  $\sigma_j$'s are simulated as: 
$$\sigma^2 \sim \text{Inv-Gamma}(\alpha=2,\beta=5).  $$ 
For each simulated data set, we find the optimal set of $k_1 = 5$ IPD studies when the $\sigma_j$'s and $\sigma_{\alpha}$'s are known without uncertainty. Next, we derive the estimates of the variance components and obtain the set of $k_1 = 5$ IPD studies using the sequential algorithm based on these estimates. Table \ref{sim_lmm_sensitivity} reports the count of exact matches between the two sets. For both the cases, we found that the selection algorithm is robust to the estimation of these parameters.

\begin{table}[!htbp]
\centering
\caption{Count of matches for a set of size $k_1=5$ out of $k=10$ studies}
\vspace*{1mm}
\begin{tabular}{|c|c|c|}
\hline
\multicolumn{1}{|l|}{\textbf{Count Match}} & {$\sigma_j^2=\sigma^2$} & {$\sigma_j^2\neq \sigma^2$}\\ 
\hline
{0} 	   & 0 	    & 0	     \\  
{1}        & 0 	    & 0      \\  
{2}        & 0 	    & 0      \\  
{3}        & 0 	    & 386      \\  
{4}        & 607 	& 3636      \\  
{5}        & 9393 	& 5974  \\ 
\hline                       
\end{tabular}
\label{sim_lmm_sensitivity}
\end{table}

\subsection{World Values Survey: Efficient Recombination}

We illustrate the significance of selection of IPD studies when combining information across studies using  data  from the World Values Survey (WVS) (\url{https://www.worldvaluessurvey.org/wvs.jsp}). The WVS is a large consortium of social survey data from around 100 regions across 6 waves (1981-1984, 1990-1994, 1995-1998, 1999-2004, 2005-2009 and 2010-2017). The data consist of scores based on questionnaire covering a broad range of topics such as economic development, democratization, religion, gender equality, social capital, and subjective well-being. 

For our analysis, we consider a simple linear mixed effects model based on the data for wave 6 in USA which contains 2232 participants. The dependent variable is life satisfaction score which ranges from 1 (completely dissatisfied) to 10 (completely satisfied). Based on financial satisfaction score, the predictor is a binary variable taking the value 1 for scores more than 5 and 0 otherwise. We consider category 1 as treatment and 0 as control for the predictor. We focus on individuals who are single and also account for heterogeneity due to age, sex, ethnicity and education in the study population through stratification. The variable age is categorized into intervals of 20 years as 0-20, 20-40, 40-60, 60-80 and more than 80. Ethnicity has 5 categories: Non-Hispanic white, Non-Hispanic Black, Hispanic, Non-Hispanic more than 2 races and other races while sex is categorized into male and female groups. The variable education has many categories starting with no formal education to doctorate degree. The strata are formed by considering different combinations of age, sex, ethnicity and education. We remove strata that have either only 1 or only 0 as response, resulting in treatment proportions to be 1 or 0. We also remove strata with number of participants (stratum size) less than 3. The total number of strata after exclusion is 30 with strata sizes ranging from 3 to 24 and the proportion of treatment ranging from 0.1 to 0.89.  Each strata will be considered as a study and meta analysis will be in combining the strata specific results to obtain overall population results.

We assume that  we have access to the AD for each study, i.e., the estimate of the treatment effect and the estimated standard error along with the number of participants in the treatment group and the control group are available for each study. The objective is to assess whether getting IPD is useful and determine which IPDs should be combined with AD to obtain maximum efficiency. We randomly sample 5 pilot studies among the 30 total studies and estimate $\sigma_{\alpha}^2=0.144$ using model (\ref{lmm_model}). First, we consider the case when $\sigma^2_j$'s are not same. The estimated variance for an all-IPD estimator of the treatment effect, $\hat{\beta}_{IPA-MA}$ is 0.033 whereas for an estimator based on all-AD estimator, $\hat{\beta}_{AD-MA}$ the  estimated variance is 0.038 resulting in an approximately $11\%$ loss in efficiency. However, if we include 5 best subset IPD studies selected using the algorithm we proposed, the estimated variance for the combined estimator, $\hat{\beta}_{IPD-AD-MA}$ is 0.035 and hence we have a $7\%$ gain in efficiency compared to $\hat{\beta}_{AD-MA}$. 

\begin{figure}
    \centering
    \includegraphics[width = 5in]{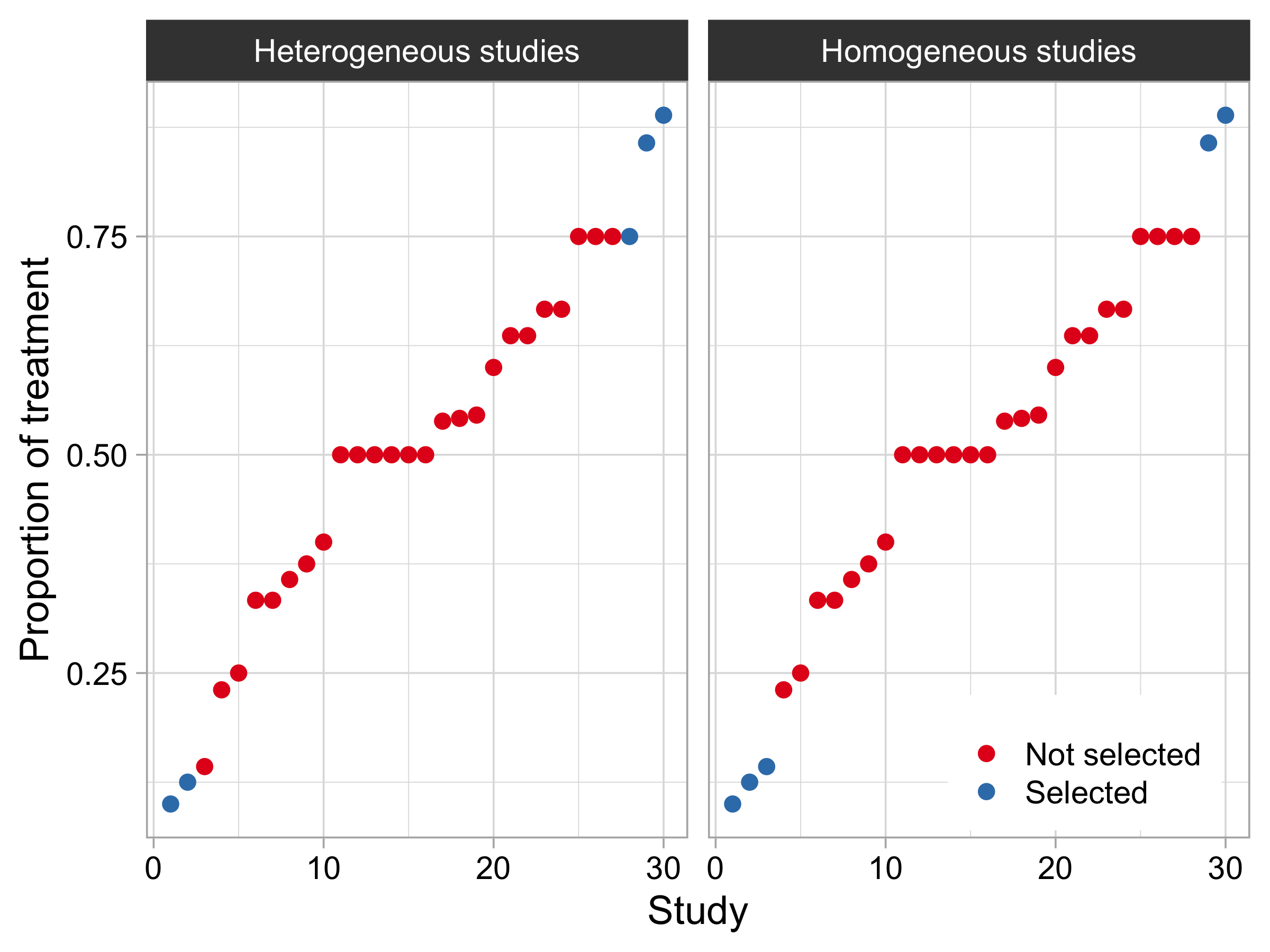}
    \caption{Plot showing the studies that are selected against the proportion of treatments when the study variances are homogeneous versus  when they are heterogeneous.}
    \label{fig5}
\end{figure}

Under the assumption $\sigma^2_j=\sigma^2$, there is an approximate $12\%$ loss in efficiency for the AD-MA estimator compared to the IPD-MA estimator. If we include the 5 best chosen IPD studies, there is a $9\%$ gain in efficiency for the IPD-AD-MA estimator  relative to the AD-MA estimator. Figure \ref{fig5} shows which 5 studies are selected using the sequential algorithm. Both assumptions yield nearly the same set of best 5 IPD studies which reinforces that the selection algorithm is not  sensitive to variance component estimation. Table \ref{rda_lmm} summarizes the pooled estimates and the standard errors for various scenarios. 

\begin{table}[!htbp]
\centering
\caption{Pooled estimate and standard error of the treatment effect for IPD-MA estimator, IPD-AD-MA estimator with 5 best subset IPD and the AD-MA estimator}
\vspace*{1mm}
\begin{tabular}{|c|c|c|c|c|c|}
\hline
\multicolumn{1}{|c|}{}           & \multicolumn{2}{c|}{$\sigma_j^2=\sigma^2$}                                        &            & \multicolumn{2}{c|}{$\sigma_j^2\neq \sigma^2$}                                    \\ \hline
\multicolumn{1}{|c|}{\textbf{IPD Trials}} & {$\boldsymbol{\hat{\beta}}$} & \textbf{se($\boldsymbol{\hat{\beta}}$)} & \textbf{IPD Trials} & {$\boldsymbol{\hat{\beta}}$} & \textbf{se($\boldsymbol{\hat{\beta}}$)} \\ \hline
None                             & 1.624                              & 0.221                                        & None                  & 1.603                              & 0.195                                   \\ \hline
1:3, 29:30                       & 1.762                              & 0.207                                        & 1:2, 28:30            & 1.746                              & 0.182                                   \\ \hline
1:30                             & 1.784                              & 0.205                                        & 1:30                  & 1.774                              & 0.179                                   \\ \hline
\end{tabular}
\label{rda_lmm}
\end{table}

\section{Aggregation in Generalized Linear Mixed Model }\label{section_glmm}

For studies with non-Gaussian data, generalized linear models (GLM) form a large class, including popular models for binary response and count response. To aggregate several GLM studies with the same parameters of interest, one could use a meta analysis approach similar to that in the linear model case. 
We consider a generalized linear mixed model (GLMM) setup which allows for random effect terms for the study effect.
We use a simple aggregation framework which will allow us to derive an efficiency expression for the treatment effect estimator similar to that in the LMM. We could then use the efficiency expression, written in terms of the treatment allocations to decide which AD could be replaced by IPD for maximum gain in efficiency. 

We present the setup in a slightly more general form with several fixed effects and random effects terms. 
Suppose, for $j =1,\dots,k_1$ studies in set $S_1$, we have access to the IPD data whereas for $k_2$ studies in set $S_2$, only the summary statistics for the parameter of interest are available where $S_1 \cup S_2= S$ is the total set of studies with $|S|=k$. 

The IPD data $(Y_{ji},X_{ji}|\beta_j,\alpha_j) \sim f_j(Y_{ji},X_{ji};\beta_j,\alpha_j)$ where $f$ is a probability density belonging to the regular exponential family. We assume a random effect model for $\beta_j$ and $\alpha_j$, where both $\beta_j$ and $\alpha_j$ may be multidimensional.  We further assume a joint multivariate gaussian model for the random effects allowing for non-zero correlation between the $\beta_j$ and $\alpha_j$ where $\beta$ is the parameter vector of interest and the vector $\alpha$ includes all nuisance parameters:
\begin{equation*}\label{eq:1}
\setlength{\jot}{10pt}
\begin{aligned}
 \begin{pmatrix} \beta_j\\
  \alpha_j\\ \end{pmatrix}\bigg\rvert \begin{pmatrix} \beta\\
  \alpha\\ \end{pmatrix}  \sim MVN\Bigg( \begin{pmatrix} \beta\\
  \alpha\\ \end{pmatrix}, \begin{pmatrix} \Sigma_{\beta\beta} & \Sigma_{\beta\alpha}\\
  \Sigma_{\alpha\beta} & \Sigma_{\alpha\alpha}\\ \end{pmatrix} \Bigg)
\end{aligned}
\end{equation*}
with $MVN$ denoting the multivariate normal density.
This is a general model and can be reduced to specific mixed effects models or fixed effects model by constraining the  corresponding components in the covariance matrix  to be zero. This setup allows having additional covariates and the parameter corresponding to the additional covariates can be included in $\alpha_j$. 

For the $k_2$ AD studies, we have the MLE estimates of the main parameter vector with their estimated covariance matrices, $(\hat{\beta}_j,\widehat{V(\hat{\beta}_j)})$. The model for random effects is given by
\begin{equation*}\label{eq:2}
\setlength{\jot}{10pt}
\begin{aligned}
\hat{\beta}_j|\beta_j &\sim N(\beta_j,\widehat{V(\hat{\beta}_j)})\\
\beta_j|\beta &\sim N(\beta,\Sigma_\beta).
\end{aligned}
\end{equation*}
We will use a composite likelihood where the likelihood from the studies with IPD and the likelihood from the studies with AD are simply multiplied together.\cite{varin2011overview} For the studies with only summary statistics available, we consider an approximate likelihood based on the asymptotic normal model for the maximum likelihood estimators (MLE). For the studies with IPD, we can write the full GLMM likelihood. But given the complexity of the GLMM likelihood involving integrals we use an approximate likelihood based on the Laplace approximation. 
We estimate the common parameter  of interest using profile likelihood approach applied to the composite likelihood. 

 The full (log) likelihood function is obtained by combining IPD studies and AD studies as:
\begin{equation*}
\setlength{\jot}{10pt}
\begin{aligned}
L(\beta,\alpha) &= \displaystyle\prod_{j \in S_1} {L_j(\beta,\alpha)} \displaystyle\prod_{j \in S_2} {L_j(\beta)},\\ 
\log L(\beta,\alpha) &= l(\beta,\alpha) = \sum_{j \in S_1}{l_j(\beta,\alpha)} + \sum_{j \in S_2}{l_j(\beta)}. 
\end{aligned}
\end{equation*}
For each $k \in S_1$, the IPD part of log-likelihood, $l_j(\beta,\alpha)$ is constructed using the Laplace approximation (where  the family $f$ satisfies the standard smoothness assumptions for the Laplace approximation to work) to expand $l_j(\beta,\alpha)$ around the MLE, $(\hat{\beta}_j,\hat{\alpha}_j)$ from study $j$ and ignoring the higher order terms $o(||\beta_j-\hat{\beta}_j||^2)$. Then, $(\beta_j,\alpha_j)$ is integrated out in the Gaussian integral to obtain
\be
\label{eq:6}
l_j(\beta,\alpha)  = l_j(\hat{\beta}_j,\hat{\alpha}_j)
            -\frac{1}{2}\Bigg[
           \begin{pmatrix}
           \hat{\beta_j} \\
           \hat{\alpha_j}\\
         \end{pmatrix} - \begin{pmatrix}
           \hat{\beta_j} \\
           \hat{\alpha_j}\\
         \end{pmatrix} \Bigg] ^T 
         \Delta_j^{-1} \Bigg[
           \begin{pmatrix}
           \hat{\beta_j} \\
           \hat{\alpha_j}\\
         \end{pmatrix} - 
         \begin{pmatrix}
           \hat{\beta_j} \\
           \hat{\alpha_j}\\
         \end{pmatrix} \Bigg]
         -\frac{1}{2}\log |\Delta_j|
\ee
where $\Delta_j = \Bigg[\begin{matrix}
         \begin{pmatrix} \Sigma_{\beta\beta} & \Sigma_{\beta\alpha}\\
  \Sigma_{\alpha\beta} & \Sigma_{\alpha\alpha}\\ \end{pmatrix} 
         + \Bigg[\begin{pmatrix}
           I_{\beta_j\beta_j} & I_{\beta_j\alpha_j}\\
           I_{\alpha_j\beta_j} & I_{\alpha_j\alpha_j}\\
         \end{pmatrix} 
         \bigg\rvert_{\hat{\beta}_j,\hat{\alpha}_j} \Bigg]^{-1}
          \end{matrix} \Bigg]^{-1}$. 
          
The AD part of likelihood for each $k \in S_2$ is given by
\begin{equation}\label{eq:7}
\setlength{\jot}{10pt}
l_j(\beta)  = {-\frac{1}{2}\begin{pmatrix}
           \hat{\beta_j} - \beta \\
         \end{pmatrix} ^T 
         \big( \widehat{V(\hat{\beta}_j)})
         + \Sigma_{\beta \beta}\big)^{-1}
         \begin{pmatrix}
           \hat{\beta_j} - \beta \\
         \end{pmatrix}}\\
         -\frac{1}{2}\log | \widehat{V(\hat{\beta}_j)})
         + \Sigma_{\beta \beta}|.
\end{equation}\\
Putting (\ref{eq:6}) and (\ref{eq:7}) together, we have the full log likelihood for $(\beta,\alpha)$ as
\beqn \label{eq:8}
l(\beta,\alpha)  &=&  \sum_{j \in S_1} l_j(\hat{\beta}_j,\hat{\alpha}_j)-\frac{1}{2}\sum_{j \in S_1}\Bigg[
           \begin{pmatrix}
           \hat{\beta_j} \\
           \hat{\alpha_j}\\
         \end{pmatrix} - \begin{pmatrix}
           \hat{\beta_j} \\
           \hat{\alpha_j}\\
         \end{pmatrix} \Bigg] ^T 
         \Delta_j^{-1} \Bigg[
           \begin{pmatrix}
           \hat{\beta_j} \\
           \hat{\alpha_j}\\
         \end{pmatrix} - 
         \begin{pmatrix}
           \hat{\beta_j} \\
           \hat{\alpha_j}\\
         \end{pmatrix} \Bigg]
         -\frac{1}{2}\sum_{j \in S_1} \log |\Delta_j| \nn \\
       &&  -\frac{1}{2}\sum_{j \in S_2} (\hat{\beta_j} - \beta)^T 
         \big(\widehat{V(\hat{\beta}_j)})
         + \Sigma_{\beta \beta}\big)^{-1}
        (\hat{\beta_j} - \beta)
         -\frac{1}{2}\sum_{j \in S_2} \log | \widehat{V(\hat{\beta}_j)}
         + \Sigma_{\beta \beta}|.
\eeqn
which then can be maximized to obtain the estimates of mean parameters and the variance components. To reduce the computational burden, we lower the dimension by expressing the mean parameter as a function of the variance components in (\ref{eq:9}).
\begin{equation}\label{eq:9}
\setlength{\jot}{10pt}
\begin{aligned}
\begin{pmatrix}
\hat{\beta}(\Sigma) \\
\hat{\alpha}(\Sigma) \\
\end{pmatrix} 
&= \Bigg[\sum_{j \in S_1} \Delta_j + \sum_{j \in S_2} \Omega_j\Bigg]^{-1} \ \Bigg[\sum_{j \in S_1} \Delta_j  \begin{pmatrix}
           \hat{\beta_j} \\
           \hat{\alpha_j}\\
         \end{pmatrix}  + \sum_{j \in S_2} \Omega_j \begin{pmatrix}
           \hat{\beta_j} \\
           \hat{\alpha_j}\\
         \end{pmatrix} \Bigg]
\end{aligned}
\end{equation}
where 
\begin{equation*}
\setlength{\jot}{10pt}
\begin{aligned}
\Delta_j &= \Delta_j(\Sigma) =
        \Bigg[
         \begin{pmatrix} \Sigma_{\beta\beta} & \Sigma_{\beta\alpha}\\
         \Sigma_{\alpha\beta} & \Sigma_{\alpha\alpha}\\ \end{pmatrix} 
         + \Bigg[\begin{pmatrix}
           I_{\beta_j\beta_j} & I_{\beta_j\alpha_j}\\
           I_{\alpha_j\beta_j} & I_{\alpha_j\alpha_j}\\
         \end{pmatrix} 
         \bigg\rvert_{\hat{\beta}_j,\hat{\alpha}_j} \Bigg]^{-1}
        \Bigg]^{-1}\\
\Omega_j &=\Omega_j(\Sigma_{\beta\beta}) =
         \Bigg[\begin{matrix}
           (\Sigma_{\beta\beta}+\widehat{V(\hat{\beta}_j)})^{-1} & 0\\
           0 & 0\\
         \end{matrix}\Bigg].
\end{aligned}
\end{equation*}
We then maximize (\ref{eq:8}) with respect of $\Sigma$ to obtain $\hat{\Sigma}$ and use the expression from (\ref{eq:9}) to obtain the combined estimator, $ \hat{\beta}(\hat{\Sigma})$ denoted by $\hat{\beta}_{IPD-AD-MA}$.
The estimated variance of $\beta$ is then given by
\begin{equation}\label{var_comb_glmm}
\setlength{\jot}{10pt}
\begin{aligned}
\widehat{V(\hat{\beta}_{IPD-AD-MA})}&=
\Bigg[\sum_{j \in S_1} \Delta_j(\hat{\Sigma}) + \sum_{j \in S_2} \Omega_j(\hat{\Sigma}_{\beta \beta})\Bigg]^{-1}_{ \ \ [1,1]}.
\end{aligned}
\end{equation} 

In the GLMM framework, for the IPD-MA estimator, we essentially combine the MLE estimates and the estimated variances for all the parameters in the model, including the parameter of interest and the nuisance parameter whereas for the AD-MA estimator, we integrate MLE estimates and the estimated variances for only the parameter of interest. Thus, there may be a loss of information for the AD-MA estimator when compared to the IPD-MA estimator depending on the correlation of the parameter of interest with the nuisance parameters. 

In GLMM, we use the data to get the MLE estimates and integrate them to derive the IPD-MA estimator. In LMM, however, one can directly use the data for the IPD studies instead of using the MLE estimates. While the general methodology of aggregation of GLMM studies can be applied to linear mixed models with the identity link, explicit computation of the likelihood and the estimators in the LMM allows more efficient strategies for data aggregation. 

\subsection{Selection in the Logistic Model}\label{glmm_selection}

For a special case of generalized linear mixed model, we provide the variance expression for IPD selection. We consider a logistic model with similar setup as (1), where the treatment effect, $\beta$ is fixed and the trial effect is random with with $\alpha_j \sim N(\alpha,\sigma^2_\alpha)$. Suppose further that, $n_{jT}$ and $n_{jC}$ be the number of persons allocated for treatment and control, respectively, for $j$th study where $n_{jT}+n_{jC}=n_j$. Let $x_{ji}$ denote the treatment allocation for individual $i$ in study $j$, with proportion of treatment $\pi_j$ for $j=1,\dots,k$ and $i=1,\dots,n_j$. The response $y_{ji}$ is binary with probability of success,
\begin{equation}\label{logistic}
\setlength{\jot}{10pt}
\begin{aligned}
P(y_{ji}=1|x_{ji}=1) &=& p_{j1} &= \frac{\exp(\alpha_j+\beta)}{1+\exp(\alpha_j+\beta)},\\
P(y_{ji}=1|x_{ji}=0) &=& p_{j0} &= \frac{\exp(\alpha_j)}{1+\exp(\alpha_j)}.
\end{aligned}
\end{equation} 
\\
Using the variance expression in (\ref{var_comb_glmm}), the variance of the combined estimator can be simplified to
\begin{equation}\label{var_comb_logistic}
\setlength{\jot}{10pt}
\begin{aligned}
V(\hat{\beta}_{IPD-AD-MA})&=
{\bigg[\sum_{j \in S_{1}} \frac{h_j}{c_j} (g_{j}-\tilde{g})^2 + \sum_{j =1}^{k} h_j^{-1}   \bigg]}^{-1}
\end{aligned}
\end{equation} 
where 
\begin{equation*}
\setlength{\jot}{10pt}
\begin{aligned}
h_j &= a_j^{-1}+b_j^{-1} = V(\hat{\beta}_j),\\
g_j &= \frac{1}{{(a_j/b_j)}^{-1}+1},\\
c_j &= \sigma^2_\alpha h_j + a_j^{-1}b_j^{-1}, \\
a_j &=  n_{jT} \ p_{j1} (1-p_{j1}),\\
b_j &=  n_{jC} \ p_{j0} (1-p_{j0}).
\end{aligned}
\end{equation*} 

The problem of using  (\ref{var_comb_logistic}) to select the ``best" combination of IPD and AD is identical to the selection problem stated in (\ref{opt_selection}). In order to use the sequential selection algorithm proposed in section \ref{selection}, the unknowns $g_j$ and $h_j/c_j$ need to be estimated using AD. There are two possible solutions.  

First, with the assumption of rare disease, $a_j/b_j = ({n_{jT}}/{n_{jC}})*(\text{odds ratio}_j)$ where the $\text{odds ratio}_j$ can be estimated from $\hat{\beta}_j$. Therefore, with this assumption  $h_j/c_j$ and $g_j$ can be computed as
\begin{equation}\label{gk_est}
\setlength{\jot}{10pt}
\begin{aligned}
g_j &= \frac{1}{({n_{jT}}/{n_{jC}}) * \exp{(\hat{\beta}_j)} + 1}\\
\frac{h_j}{c_j} &=  \frac{1/\widehat{V(\hat{\beta}_j)}}{\sigma^2_\alpha/\widehat{V(\hat{\beta}_j)})+g_j(1-g_j)}
\end{aligned}
\end{equation} 
provided we have AD for all studies along with $n_{jT}$ and $n_{jC}$. We can then compute the variance expression (\ref{var_comb_logistic}) to choose the IPD studies that `maximize' the efficiency of the combined estimator using the sequential algorithm (\ref{seq_algm}) .

The second solution requires knowledge of the number of cases and controls for each study in addition to treatment and control group totals. In fact, with this information we can reconstruct the individual $2 \times 2$ contingency tables and hence estimate $g_{j}$ and $h_{j}/c_j$ even without the rare disease assumption. However, efficient selection of IPD studies is of no additional advantage for binary data in RCTs in case the $2 \times 2$ contingency tables are available since one can construct the IPD from the table. 

\subsection{Simulation results}

We illustrate the advantages of  combining estimates in GLMM through a limited simulation experiment using binary data and the  logistic link.  Suppose $X_{ji}$ denotes the treatment allocation for individual $i$ in study $j$, with proportion of treatment, $\pi_j$ where $j=1,\dots,k$ and $i=1,\dots,n_j$. The response $y_{ji}$ is binary with probability of success as in model (\ref{logistic}) where both $\beta_j$ and $\alpha_j$ are assumed to be random with $\beta_j|\beta \sim N(\beta,\Sigma_\beta)$ and $\alpha_j|\alpha \sim N(\alpha,\Sigma_\alpha)$, independently, where the true parameters are taken to be $\beta=0.5,\alpha=0.5,\sigma^2_\beta=0.5$ and $\sigma^2_\alpha=0.5$. We simulated the true $\pi_j$'s from a uniform distribution over the interval $[0,1]$. The results are shown in Table \ref{table1}. We present the estimates, estimated bias and standard error for different choices of $n=n_j$ and $k$ along with the relative efficiency with respect to the IPD-MA estimator. Similar to the linear model, we could have different combinations of IPD and AD for the IPD-AD-MA estimator, even when the percentage of IPD is fixed. However, since there can be many such combinations for $k=50$, we only report the results for one randomly chosen combination for each possibility of no IPD, $20\%$ IPD, $40\%$ IPD, $60\%$ IPD, $80\%$ IPD and all IPD.

The mean square error (MSE) for the all AD estimator is 0.021 for $k=50,n=100$, 0.013 for $k=50,n=500$, 0.012 for $k=100,n=100$ and 0.006 for $k=100,n=500$. The MSE for the all IPD estimator is 0.019 when $k=50,n=100$ and 0.010 for $k=100,n=100$ and stays the same for other choices of $k$ and $n$. While the bias doesn't change much when $n$ is fixed and $k$ is varying, the MSE decreases with increase in either sample size or number of studies or both. On the other hand, the relative efficiency approaches 1 as percentage of IPD studies is increased for each choice of $k$ and $n$.

\begin{table}[!htbp]
\centering
\caption{Model parameter estimates for different choices of $n$ and $k$}
\vspace*{1mm}
\begin{tabular}{|c|c|c|c|c|c|}
\hline
\multicolumn{1}{|l|}{\textbf{Scenario}} & \textbf{(\% IPD, \% AD)} & \textbf{Estimate} & \textbf{Bias} & \textbf{Std. Error} & \textbf{Rel. eff.} \\ 
\hline
{k=50, n=100}   & (0, 100)                                  & 0.43                               & -0.07                          & 0.127                                        & 0.926                                         \\
                               & (20, 80)                                  & 0.43                               & -0.07                          & 0.127                                        & 0.925                                         \\
                               & (40, 60)                                  & 0.431                              & -0.069                         & 0.126                                        & 0.94                                          \\
                               & (60, 40)                                  & 0.433                              & -0.067                         & 0.125                                        & 0.961                                         \\
                               & (80, 20)                                  & 0.436                              & -0.064                         & 0.125                                        & 0.982                                         \\
                               & (100, 0)                                  & 0.44                               & -0.06                          & 0.126                                        & 1.000                                         \\ 
                               \hline
{k=50, n=500}   & (0, 100)                                  & 0.482                              & -0.018                         & 0.113                                        & 0.975                                         \\
                               & (20, 80)                                  & 0.482                              & -0.018                         & 0.113                                        & 0.976                                         \\
                               & (40, 60)                                  & 0.483                              & -0.017                         & 0.113                                        & 0.979                                         \\
                               & (60, 40)                                  & 0.483                              & -0.017                         & 0.113                                        & 0.983                                         \\
                               & (80, 20)                                  & 0.485                              & -0.015                         & 0.113                                        & 0.986                                         \\
                               & (100, 0)                                  & 0.488                              & -0.012                         & 0.112                                        & 1.000                                         \\ 
                               \hline
{k=100, n=100}  & (0, 100)                                  & 0.438                              & -0.062                         & 0.09                                         & 0.875                                         \\
                               & (20, 80)                                  & 0.438                              & -0.062                         & 0.09                                         & 0.873                                         \\
                               & (40, 60)                                  & 0.439                              & -0.061                         & 0.09                                         & 0.89                                          \\
                               & (60, 40)                                  & 0.441                              & -0.059                         & 0.089                                        & 0.911                                         \\
                               & (80, 20)                                  & 0.444                              & -0.056                         & 0.089                                        & 0.942                                         \\
                               & (100, 0)                                  & 0.448                              & -0.052                         & 0.088                                        & 1.000                                         \\ 
                               \hline
{k=100, n=500}  & (0, 100)                                  & 0.482                              & -0.018                         & 0.074                                        & 0.973                                         \\
                               & (20, 80)                                  & 0.482                              & -0.018                         & 0.074                                        & 0.975                                         \\
                               & (40, 60)                                  & 0.483                              & -0.017                         & 0.074                                        & 0.979                                         \\
                               & (60, 40)                                  & 0.484                              & -0.016                         & 0.074                                        & 0.986                                         \\
                               & (80, 20)                                  & 0.485                              & -0.015                         & 0.074                                        & 0.994                                         \\
                               & (100, 0)                                  & 0.488                              & -0.012                         & 0.074                                        & 1.000                                         \\ 
\hline
\end{tabular}
\label{table1}
\end{table}

\subsection{Real data analysis}

We consider the dataset in \textit{Yusuf et al. (1985)} which includes results from the long-term trials of oral beta blockers on its effectiveness for reducing mortality.\cite{yusuf1985beta} The data consists of $2\times2$ tables from 22 clinical trials. We illustrate the aggregation of (log) odds ratio under the logistic model (\ref{logistic}) with independent random effects for the study effect and treatment effect. Although the full data is available, for the purpose of illustration we assume access to IPD for some trials and access to only meta-analysis results for the remaining trials. The cohort size for the smallest trial is 77 whereas the largest trial has a cohort of 3837 individuals. To show the significance of selection for the logistic model, we randomly sampled $n_j=50$ individuals for each study $j$.

We assume  we have access to the AD for each study,i.e. the MLE estimates of the log-odds ratio and its standard error along with the treatment and control group totals. We apply the sequential algorithm using the estimates in (\ref{gk_est}) where the $\sigma^2_{\alpha}$ is estimated using a randomly selected pilot IPD studies. In practice, it can be estimated using IPD studies that are conveniently accessible since the $\sigma^2_{\alpha}$ seems to have little or no effect in choosing the final set of selected IPD studies. The estimated variance for the IPD-MA estimator and AD-MA estimator are  0.033 and 0.044, respectively, which implies a $26\%$ loss in efficiency. If we include the best 5 IPD studies chosen using the sequential algorithm, the estimated variance for the IPD-AD-MA estimator is 0.035, resulting in a mere $6\%$ loss in efficiency relative to the IPD-MA estimator, which is a gain of $26\%$ efficiency relative to the AD-MA estimator. However, the IPD-AD-MA estimator with the worst 5 IPD studies results in a estimated variance of 0.044, which is no any efficiency gain from the AD-MA estimator.

The pooled estimate (log) odds ratio with its standard error for the IPD-MA estimator, the best combination of IPD-AD studies with $k_1=2,5$ and $8$, respectively and the AD-MA estimator are reported in Table \ref{rda_glmm}. The IPD trials in the table are ordered with respect to the $g_j$, analogous to the proportion of treatment, $\pi_j$ in the linear mixed effects model. The standard error for the best combined estimator decreases as number of IPD studies is increased. In conclusion, a combined estimator with just 8 selected IPD studies out of the total 22 studies can achieve efficiency very close to the IPD-MA estimator.

\begin{table}[!htbp]
\centering
\caption{Pooled estimate and standard error of the (log) odds ratio for the IPD-MA estimator, the best IPD-AD-MA estimator with 2 IPDs, 5 IPDs and 8 IPDs, respectively and the AD-MA estimator}
\vspace*{1mm}
\begin{tabular}{|c|c|c|}
\hline
\multicolumn{1}{|l|}{\textbf{IPD Trials}} & {$\boldsymbol{\hat{\beta}}$} & \textbf{se($\boldsymbol{\hat{\beta}}$)} \\ 
\hline
None                             & 0.001                              & 0.211                                        \\ \hline
1,22                            & 0.006                              & 0.206                                        \\ \hline
1:3,21:22                   & 0.017                              & 0.204                                        \\ \hline
1:5,20:22                    & 0.001                              & 0.200                                        \\ \hline
1:22                             & 0.057                              & 0.199                                        \\ \hline
\end{tabular}
\label{rda_glmm}
\end{table}

\section{Discussion}

In this paper, we provide a method of combining information across independent studies for a generalized mixed effects model in a multivariate setup. We show that the combined estimator is efficient compared to the all IPD estimator asymptotically through simulation. For a much simpler linear mixed effects model with heterogeneous studies, we investigate the performance of the combined estimator for various distribution of treatment proportion across studies. We advocate a method for selection of AD studies over IPD studies to ensure a fully efficient combined estimator. When the number of studies is large, we propose an approximate sequential algorithm to select the best combination of IPD and AD studies.

We assume the same set of covariates for each study. In the  future, we plan to expand our method to include disparate covariate information for multivariate mixed effects models. Another possible direction would be to quantify the degree of unbalancedness of proportion of treatments among studies. That would help in investigating the impact of selecting AD studies for multiple treatment problems as well. It would be interesting to extend the sequential algorithm to a more general GLM framework, which requires a result on optimality of the AD estimators in terms of the design parameters. Once such a result has been established then an analogous algorithm could be devised by optimizing a measure of departure from the optimality condition.

We also look at an interesting application of our novel selection algorithm in the context of the recently popular `Split and Conquer' approach to analysis of `Big Data'. With the increasing need of resources to store and analyze large data, meta-analysis methods are becoming very popular. Due to the availability and accessibility of massive amount of data in several fields, a new research paradigm is focused towards divide and recombine (D\&R) approach for big data analysis.\cite{lee2017sufficiency,chen2014split,cheung2016analyzing}  The main goal of our method is to aid in the analysis of Big Data like Genome Wide Association Studies, Electronic Health Records, large socio-economic survey data,  using the split, analyze and aggregate approach in order to reduce computational and storage limitations but still provide estimates with efficiency close to that from the IPD analysis. 

\section{Data Availability Statement}

The data that support the findings of this study are available in github repository 'meta-analysis' at https://github.com/agneha2010/meta-analysis. These data were derived from the following resources available in the public domain: \textit{Yusuf et al. (1985)} and https://www.worldvaluessurvey.org/wvs.jsp.\cite{yusuf1985beta} 

\bibliographystyle{WileyNJD-AMA}
\bibliography{main_reference.bib}

\end{document}